\documentstyle[12pt,epsfig]{article}
\def\lapprox{\lower .7ex\hbox{$\;\stackrel{\textstyle <}{\sim}\;$}}
\def\gapprox{\lower .7ex\hbox{$\;\stackrel{\textstyle >}{\sim}\;$}}

\def\e{\epsilon}
\def\d{{\rm d}}

\def\barq{\bar{q}}

\def\xa{x_{1}}
\def\xb{x_{2}}
\def\xaa{x_{1}^{0}}
\def\xbb{x_{2}^{0}}

\def\Li{{\rm Li}}

\begin{document}
\begin{titlepage}
\vspace*{-1cm}
\begin{flushright}
DESY--97--014   \\
February 1997 \\
\end{flushright}                                
\vskip 3.5cm
\begin{center}
{\Large\bf QCD corrections to the longitudinally polarized}
\vskip .1cm
{\Large\bf Drell--Yan process}
\vskip 1.cm
{\large  T.~Gehrmann} 
\vskip .4cm
{\it DESY, Theory Group, D-22603 Hamburg, Germany}
\end{center}
\vskip 3.cm

\begin{abstract}
In this paper we 
calculate the ${\cal O}(\alpha_s)$ corrections to the $x_F$- and 
$y$-distributions of lepton pairs produced in collisions of longitudinally 
polarized hadrons. The numerical importance of these corrections is studied 
and consequences for the extraction of the polarized sea quark distributions 
from a measurement of the longitudinally polarized Drell--Yan cross section 
are discussed. 
\end{abstract}


\vfill
\end{titlepage}                                                                
\newpage                                                                       

\section{Introduction}
The production of lepton pairs in hadron collisions, the Drell--Yan 
process~\cite{drellyan}, is one of the most powerful tools to probe the 
structure of hadrons. Its parton model interpretation is straightforward 
--~the process is induced by the annihilation of a quark--antiquark
pair into a virtual photon which subsequently decays into a lepton pair. 
The Drell--Yan process in proton--proton or proton--nucleus collisions
therefore provides a direct probe of the antiquark densities in protons 
and nuclei. 

The naive quark parton model predicts scaling of the Drell--Yan cross 
section in the variable $\tau\equiv M^2/S$, where $M$ is the invariant 
mass of the lepton pair and $\sqrt{S}$ 
is the hadron--hadron centre-of-mass energy.
Apart from the invariant mass distribution 
$\d \sigma/\d M^2$ 
one usually studies the distribution of the lepton pairs as function 
of the Feynman-parameter 
\begin{equation}
x_F \equiv \frac{2q_z}{\sqrt{S}} 
\end{equation}
or of the hadron--hadron centre-of-mass rapidity
\begin{equation} 
y \equiv \frac{1}{2}\, \ln\frac{q_0+q_z}{q_0-q_z}\; ,
\end{equation}
where $q$ denotes the four-momentum of the lepton pair in the hadron--hadron 
centre-of-mass system.
The resulting distributions 
\begin{displaymath}
\frac{\d \sigma}{\d M^2 \d x_F} \qquad \mbox {and} \qquad
\frac{\d \sigma}{\d M^2 \d y} 
\end{displaymath}
can be directly related to the $x$-dependence of the parton distribution 
functions in beam and target. 
Moreover, most fixed target experiments have only a limited 
kinematic coverage in $x_F$ or $y$, such that only these distributions 
can be measured without extrapolation into experimentally inaccessible 
regions.

A summary of the experimental data on the Drell--Yan process in unpolarized 
$pp$-, $pN$-, $\bar{p}N$- and $\pi N$-collisions can be found in~\cite{dydata}.
These data are used to determine the large-$x$ behaviour of
antiquark distributions in the proton (e.g.~\cite{mrsap,grv})
and to study the parton 
structure of the pion~\cite{smrs}. 
In recent times, the difference between Drell--Yan cross sections in 
$pp$ and $pd$ collisions~\cite{seaasy}
has provided valuable information on the
asymmetry between the $\bar{u}$- and $\bar{d}$-quark distributions in the 
proton~\cite{na51}. 

Up to now, the only experimental 
information on the polarized parton distributions in the proton
comes from deep inelastic lepton--nucleon scattering 
experiments~\cite{g1exp}, measuring the polarized
structure function $g_1(x,Q^2)$. This structure function probes one
particular combination of the polarized quark distributions, its mere 
knowledge is therefore insufficient for a determination of all different
quark and antiquark distributions. This lack of knowledge is reflected in
the results of recent fits of polarized parton distribution 
functions~\cite{gs,grsv,bfr}. While the polarized valence quark distributions
can be determined with some precision from the experimental data, only little
information can be extracted on the total magnitude of the polarized sea quark 
distribution; the flavour structure of the polarized quark sea is at present
completely undetermined.

Experimental information on the polarized antiquark distributions can be 
obtained from several observables. Charged current exchange in polarized 
electron--proton scattering can probe the polarized weak structure 
functions~\cite{weak}, which are sensitive to a different combination of
quark distributions than their electromagnetic counterparts. Combining
electromagnetic and weak structure functions in a global parton distribution
fit, it should in principle be possible to disentangle the different 
quark and antiquark flavours. A direct probe of the polarized sea quark
distributions would be the Drell--Yan process in polarized proton--nucleon
collisions~\cite{dypheno}. 
Such a measurement could be feasible at RHIC~\cite{rhic} or at 
HERA~\cite{HERA}.

The rather large QCD corrections to the {\it unpolarized} Drell--Yan process
suggest, that a reliable interpretation of the Drell--Yan cross 
section in terms of partonic distribution functions is only possible if 
QCD corrections are included. This paper aims to derive the ${\cal O}
(\alpha_s)$ corrections to the $x_F$- and $y$-distributions of lepton
pairs in longitudinally
polarized hadron--hadron collisions and to study the impact of 
these corrections on the Drell--Yan cross section.

For the unpolarized Drell--Yan process, QCD corrections have been 
calculated up to ${\cal O}(\alpha_s^2)$ for the invariant mass 
distribution~\cite{dym1,dym2} and up to ${\cal O}(\alpha_s)$ for the 
rapidity and $x_F$-distributions~\cite{aem,kubar}. For the latter,
parts of the ${\cal O}(\alpha_s^2)$ corrections are known~\cite{dydxf2}. 
Inclusion of the ${\cal O}(\alpha_s)$--corrections changes the Drell--Yan 
cross section at fixed target energies by up to 50\%, the 
${\cal O}(\alpha_s^2)$ corrections amount up to 10\%. 
At small transverse momenta of the lepton pair, the QCD corrections are
dominated by the emission of soft gluons. The effects of soft gluon emission
at arbitrary order in perturbation theory have been studied for the 
Drell--Yan process in~\cite{dysoft}, yielding a resummation of 
the dominant corrections to all orders. 

All these results can in principle be applied to the production
of vector bosons at hadron colliders. However, in the case of $W$--boson 
production, one has to take into
account that $W$--bosons in hadronic collisions can be identified only in 
their $l\nu_l$ decay mode, where just the lepton is observed. All above 
calculations cannot make predictions for the kinematic distributions 
of the vector boson decay products in the hadron--hadron centre-of-mass frame.
Separate calculations of the next-to-leading order QCD corrections 
to the transverse mass distribution of the $W$--boson~\cite{wtrans}, to the 
polarization density matrix of its decay products~\cite{wdecay} 
 and to the 
production rates of $W$+0,1 jets~\cite{wjets}
in unpolarized hadronic collisions have been performed. 

QCD corrections to the Drell--Yan process in polarized hadron--hadron
collisions have been derived up to ${\cal O}(\alpha_s)$
for the invariant mass distribution in 
longitudinally~\cite{ratcliffe,weber,kamal} and 
transversely~\cite{dyperp} polarized collisions. The 
${\cal O}(\alpha_s)$ corrections to the 
rapidity distributions are known for longitudinally~\cite{weber}
and transversely~\cite{dyperp} polarized collisions. Furthermore, a 
resummation of the effects of soft gluons in longitudinally polarized 
collisions has been performed in~\cite{weber}. 

A fully consistent study of the Drell--Yan process at next-to-leading
order was until now only possible in the unpolarized case, as the polarized 
parton distributions could only be determined at leading accuracy. With the 
recently calculated polarized two--loop splitting functions~\cite{nlosplit},
the polarized distributions can now be determined to next-to-leading order
from fits~\cite{gs,grsv,bfr} to polarized structure function data. Having a 
complete calculation of the two-loop splitting functions available, it is 
now furthermore possible to define consistent scheme
transformation prescriptions~\cite{nlosplit,bfr} for parton
distributions, splitting functions and parton level cross sections at 
next-to-leading order. The scheme dependence of the parton level cross 
sections for polarized deep inelastic scattering has lead to some 
discussion~\cite{g1disc} about the partonic interpretation of 
measurements~\cite{g1exp} of polarized lepton--nucleon scattering.
The scheme dependence of the QCD corrections to the polarized Drell--Yan 
process has been addressed in~\cite{weber,kamal}.

This paper is organized as follows:
in Section~\ref{sec:calc}, we derive the ${\cal O}(\alpha_s)$--corrections
to the $x_F$-distribution of Drell--Yan lepton pairs produced
in longitudinally polarized hadron--hadron collisions. We discuss the 
representation of $\gamma_5$ used in our calculation and motivate our
choice of factorization scheme. Furthermore,
a brief rederivation of the corrections to 
the invariant mass and $y$-distributions is given.
We numerically 
estimate the importance of these corrections at fixed target energies in 
Section~\ref{sec:num}. Section~\ref{sec:conc} contains a brief 
summary of our results and concluding remarks. In the Appendix, we collect 
several identities used in the calculation.

\section{Perturbative corrections to the polarized Drell-Yan process}
\label{sec:calc}
The QCD corrections to the $x_F$-distributions in the unpolarized Drell--Yan
process were originally derived using dimensional regularization
by Altarelli, Ellis and Martinelli~\cite{aem}. In a later publication, 
Kubar {\it et al.}~\cite{kubar} confirmed the results of~\cite{aem}
 and derived the QCD
corrections to the $y$-distributions. The calculation of Kubar {\it et al.} 
was performed using a non--zero gluon mass as regulator. The results obtained 
for the $x_F$-distributions with this method were considerably simpler than
the original results of~\cite{aem}. In the calculation of the QCD corrections
to the $x_F$- and $y$-distributions in the 
polarized Drell--Yan process, we shall work in dimensional 
regularization and follow closely the method of~\cite{aem}. Using several
identities listed in the Appendix, we are able to simplify our results into
a form similar to the unpolarized results of~\cite{kubar}. 
In the following, we give an explicit 
derivation of the corrections to the $x_F$-distribution.
Furthermore, we briefly summarize
the results of a rederivation of QCD corrections to
the $y$-distribution within the same method.
These corrections were originally obtained by Weber~\cite{weber}
as a 'by-product' in the calculation of the soft gluon resummation in the 
polarized Drell--Yan process.   

Disregarding correlations between the final state 
lepton plane and the direction of the incoming hadrons, it is sufficient to 
consider the production of an off--shell photon of invariant mass 
$M^2$, the polarization sum of the outgoing photon can be taken to 
be $-g^{\mu\nu}$. The lepton pair production cross section can be retrieved 
by multiplying the results with a factor $\alpha/(3\pi M^2)\d M^2$ for the 
decay $\gamma^* \to l^+l^-$. 

\begin{figure}[t]
\begin{center}
~ \epsfig{file=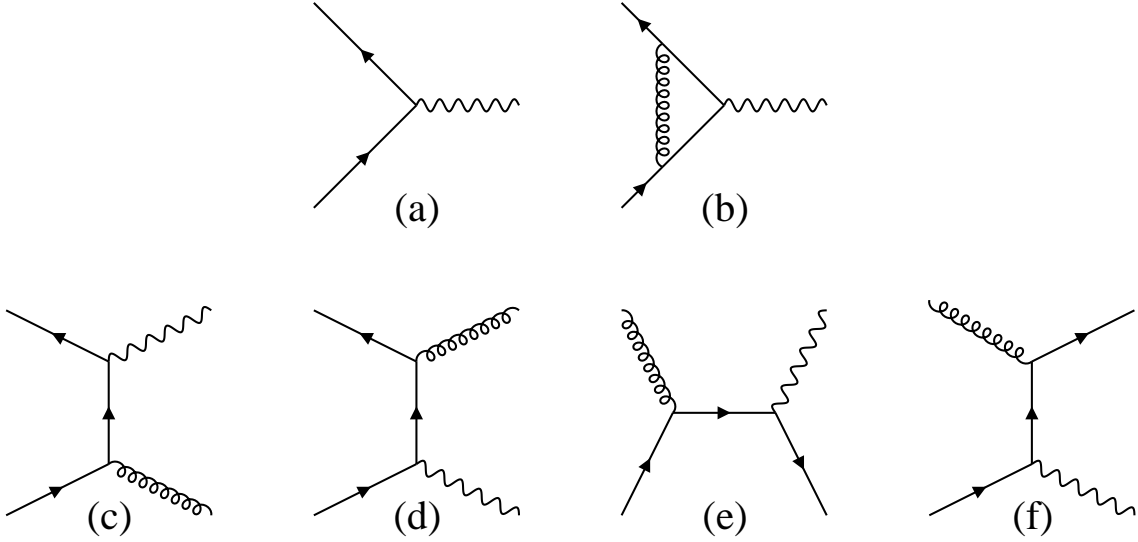,width=15cm}
\caption{Parton level subprocesses contributing to the Drell--Yan process up 
to ${\cal O}(\alpha_s)$: lowest order $q\barq$--annihilation (a); virtual (b) 
and real (c,d) gluon corrections to $q\barq$--annihilation; quark--gluon
Compton scattering (e,f).}
\label{fig:feynman}
\end{center}
\end{figure}

Only the lowest order contribution to the Drell--Yan process 
(Fig.~\ref{fig:feynman}(a)) yields a finite cross section. All higher order 
parton level cross sections contain infinities. 
To make these infinities explicit,
we work in dimensional regularization with $d=4-2\e$ dimensions. This treatment
faces the problem of describing $\gamma_5$ in $n\neq 4$. We shall use the 
$\gamma_5$--prescription~\cite{HVBM} of 't~Hooft, Veltman, Breitenlohner 
and Maison (HVBM), which has been consistently 
used in the derivation of the next-to-leading
order corrections to the polarized splitting functions~\cite{nlosplit}.
In this prescription, one restricts the anticommutation property    
of $\gamma_5$ to the physical four dimensions, while $\gamma_5$ commutes in
the remaining $n-4$ dimensions. The major drawback of this formalism in the 
$\overline{{\rm MS}}$--scheme is the non--conservation of the 
flavour non--singlet axial vector currents~\cite{laring5} 
due to a non--vanishing first 
moment of the corresponding non--singlet NLO splitting function $\Delta
P_{qq,+}$. To restore the conservation of this non--singlet axial vector
current, a further scheme transformation~\cite{laring5}
 of the results obtained in the
HVBM formalism is needed~\cite{nlosplit}\footnote{This scheme transformation 
was first considered by Weber~\cite{weber} to restore the conservation of
quark helicity in the $q\barq$--contribution to the polarized Drell--Yan 
process.}. We have to apply the same scheme transformation to our results 
for the parton-level Drell--Yan cross sections at ${\cal O}(\alpha_s)$. 

We define the hadron cross sections:
\begin{equation}
\d \Delta \sigma \equiv \frac{1}{2} \left( \d \sigma^{++} - \d \sigma^{+-} 
\right), \qquad \d \sigma \equiv \frac{1}{2} 
\left( \d \sigma^{++} + \d \sigma^{+-} \right),
\end{equation}
where $(+)$ and $(-)$ denote positive and negative hadron helicities.
In the parton model picture, these cross sections can be 
expressed as convolutions of (polarized)
parton distributions $[\Delta]f_i$ with parton subprocess 
cross sections $[\Delta]\hat{\sigma}_{ij}$:
\begin{equation}
\d [\Delta] \sigma = \int \d \xa \d \xb 
\sum_{ij} \d [\Delta] \hat{\sigma}_{ij}(\xa,\xb,\mu_F^2) \,[\Delta] 
f_i(\xa,\mu_F^2) \,[\Delta] f_j(\xb,\mu_F^2).
\end{equation}
We have defined 
\begin{equation}
\d \Delta \hat{\sigma} \equiv \frac{1}{2} 
\left( \d \hat{\sigma}^{++} - \d \hat{\sigma}^{+-} 
\right), \qquad \d \hat{\sigma} \equiv \frac{1}{2} 
\left( \d \hat{\sigma}^{++} + \d \hat{\sigma}^{+-} \right),
\end{equation}
where $(+)$ and $(-)$ now denote positive and negative parton helicities.

The lowest order bare cross section for the $q\barq\to\gamma^*$ contribution 
(Fig.~\ref{fig:feynman}.(a)) reads in $d=4-2\e$ dimensions:
\begin{equation}
\Delta \hat{\sigma}^{0} = - \frac{4 \pi^2 \alpha e_q^2}{3M^2} (1+\e) 
\delta (1-z),
\end{equation}
where we define $z=M^2/s$, with 
$\sqrt{s}$ 
being the parton--parton centre-of-mass energy. The factor $(1+\e)$ is 
common to all contributions and forms part of the overall normalization 
of the cross section. 

The virtual gluon correction (Fig.~\ref{fig:feynman}.(b)) to this process 
yields
\begin{equation}
\Delta \hat{\sigma}^{1,V}  
= - \frac{4 \pi^2 \alpha e_q^2}{3M^2} (1+\e) 
\delta (1-z) \; \frac{\alpha_s}{2\pi} C_F \left(\frac{4\pi\mu^2}
{M^2}\right)^{\e} \frac{1}{\Gamma (1-\e)} \left( -\frac{2}{\e^2}-\frac{3}{\e}
-8 +\pi^2\right).
\label{eq:qqvirt} 
\end{equation}
The correction due to real gluon emission (Fig.~\ref{fig:feynman}.(c,d)) takes
the form: 
\begin{eqnarray}
\Delta \hat{\sigma}^{1,R} & = &
- \frac{4 \pi^2 \alpha e_q^2}{3M^2} (1+\e)  \; 
\frac{\alpha_s}{2\pi}C_F \left(\frac{4\pi\mu^2} 
{M^2}\right)^{\e} \frac{1}{\Gamma (1-\e)} z^{1+\e}(1-z)^{1-2\e} \nonumber \\
& & \hspace{-0.7cm} \times \int_0^1 \d z_1 z_1^{-\e}(1-z_1)^{-\e}
\left[ (1+3\e)\left(\frac{1-z_1}{z_1}+\frac{z_1}{1-z_1}\right) + 
\frac{2z}{(1-z)^2z_1(1-z_1)} \right]\!,
\label{eq:qqreal}
\end{eqnarray}
where we have replaced the parton--parton scattering
angle $\Theta^*$ by $z_1\equiv(1+\cos\Theta^*)/2$. Terms proportional to $\e$ 
have been omitted if they do not yield a finite contribution to the 
final result. 
Finally, the contribution from the quark--gluon Compton scattering process 
(Fig.~\ref{fig:feynman}.(e,f)) reads:
\begin{eqnarray}
\Delta \hat{\sigma}^{1,C} & = &
- \frac{4 \pi^2 \alpha e_q^2}{3M^2} (1+\e)  \; 
\frac{\alpha_s}{2\pi}T_F \left(\frac{4\pi\mu^2} 
{M^2}\right)^{\e} \frac{1}{\Gamma (1-\e)} z^{1+\e}(1-z)^{1-2\e} \nonumber \\
& & \hspace{-0.7cm} \times \int_0^1 \d z_1 z_1^{-\e}(1-z_1)^{-\e}
\left[ \frac{1}{(1-z)z_1} +2 - (1+\e) \frac{2}{z_1} -(2-z_1)(1-z)
 \right] . 
\label{eq:qgreal}
\end{eqnarray}
Using eqs.~(\ref{eq:plusid}), one can make the poles in $\e$ in the 
annihilation and Compton contributions explicit:
\begin{eqnarray}
\Delta \hat{\sigma}^{1,A} & \equiv &
\Delta \hat{\sigma}^{1,V} + \Delta \hat{\sigma}^{1,R} \nonumber \\
& = & - \frac{4 \pi^2 \alpha e_q^2}{3M^2} (1+\e)  \; 
\frac{\alpha_s}{2\pi}C_F \left(\frac{4\pi\mu^2} 
{M^2}\right)^{\e} \frac{1}{\Gamma (1-\e)} \nonumber \\
&& \hspace{-0.6cm}
\times \int_0^1 \d z_1\, z\Bigg\{\left(\frac{\pi^2}{2}-4\right) \delta (1-z)
\left(\delta(z_1) + \delta (1-z_1)\right) \nonumber \\
& & \hspace{-0.2cm} + \delta(1-z) \Bigg[-\frac{1}{\e}\left( 
\frac{1}{(z_1)_{+}} 
+ \frac{1}{(1-z_1)_{+}}\right) + \left(\frac{\ln z_1}{z_1}\right)_{+} 
+ \left(\frac{\ln (1-z_1)}{1-z_1}\right)_{+} \nonumber \\
& & \hspace{0.8cm}
+ \frac{\ln z_1}{1-z_1}
+ \frac{\ln(1-z_1)}{z_1} \Bigg] \nonumber \\
&  & \hspace{-0.2cm}
+ \left[\delta(z_1)+\delta(1-z_1)\right]\, \Bigg[ -\frac{1}{\e} \left[
\frac{2}{(1-z)_{+}} -1-z + \frac{3}{2} \delta (1-z)\right] \nonumber \\
& & \hspace{0.1cm} + (1-z) \left[\ln \frac{(1-z)^2}{z} -3 \right] -2 
\ln \frac{(1-z)^2}{z} + 4 \left(\frac{\ln (1-z)}{1-z}\right)_{+} 
-\frac{2}{1-z} \ln z\Bigg] \nonumber \\
& & \hspace{-0.2cm} + \left[ \frac{1}{(z_1)_{+}} + \frac{1}{(1-z_1)_{+}} 
\right] \,\left[ -1-z+ \frac{2}{(1-z)_{+}}\right] -2(1-z)\Bigg\}, 
\label{eq:qqbare}\\
\Delta \hat{\sigma}^{1,C} & = & 
 - \frac{4 \pi^2 \alpha e_q^2}{3M^2} (1+\e)  \; 
\frac{\alpha_s}{2\pi}T_F \left(\frac{4\pi\mu^2} 
{M^2}\right)^{\e} \frac{1}{\Gamma (1-\e)} \nonumber \\
& & \hspace{-0.6cm}
\times \int_0^1 \d z_1\, z\Bigg\{ -\frac{1}{\e} \delta(z_1) \left[ 2z-1\right]
+ \delta(z_1) \left[ (2z-1) \ln \frac{(1-z)^2}{z} + 2-2z \right] 
\nonumber\\ 
& & \hspace{-0.2cm} + \frac{1}{(z_1)_{+}}  \left[ 2z-1\right]
+ 2 (1-z) - (2-z_1) (1-z)^2 \Bigg\}  .
\label{eq:qgbare}
\end{eqnarray}
The $x_F$-differential cross section can now be obtained by multiplying the 
above expressions by
\begin{displaymath}
\int \d x_F \,\delta\left(x_F-\left[z_1(\xa+\xb)(1-z)+z\xa-\xb \right]\right) .
\end{displaymath}
Using the definitions 
\begin{displaymath}
x_F = \xaa-\xbb \qquad \mbox{and} \qquad \tau = \xaa\xbb ,
\end{displaymath}
one can fix 
\begin{equation}
\xaa = \frac{1}{2} \left( x_F + \sqrt{x_F^2+4\tau} \right), \qquad
\xbb = \frac{1}{2} \left( -x_F + \sqrt{x_F^2+4\tau} \right)
\label{eq:xfdef}
\end{equation}
for the $x_F$-distributions. 
The parton level cross sections 
$\Delta \hat{\sigma}^{1,A}$ and $\Delta \hat{\sigma}^{1,C}$ can then be 
rewritten using 
eqs.~(\ref{eq:delx1})--(\ref{eq:delx4}).

The resulting parton level cross sections 
still contain divergences associated with collinear 
singularities in the initial state. These collinear singularities are 
process independent and are absorbed into the bare parton distribution 
functions by a mass factorization procedure. The only distribution relevant 
in our case is the polarized quark distribution. Performing the mass 
factorization in the $\overline{{\rm MS}}$--scheme and making a scheme 
transformation~\cite{nlosplit,weber} 
to remove spurious terms due to the HVBM prescription for $\gamma_5$ 
yields for the bare quark distribution up to ${\cal O}(\alpha_s)$:   
\begin{eqnarray}
\Delta q(x) &=& 
\Delta q(x,\mu_F^2) - \frac{\alpha_s}{2\pi} \int_x^1 \frac{\d x'}{x'} 
\left[  -\frac{1}{\e} \frac{1}{\Gamma(1-\e)} 
\left(\frac{4\pi\mu^2}{\mu_F^2}\right)^{\e} \Delta P_{qq}(x') + 
\Delta z_{qq} (x') 
\right] \Delta q(x/x') \nonumber \\
& & -  \frac{\alpha_s}{2\pi} \int_x^1 \frac{\d x'}{x'}
\left[ -\frac{1}{\e} \frac{1}{\Gamma(1-\e)} 
\left(\frac{4\pi\mu^2}{\mu_F^2}\right)^{\e} \Delta P_{qg} (x') \right] \Delta 
G(x/x') ,\label{eq:mfac}
\end{eqnarray}
with~\cite{ap}:
\begin{eqnarray*}
\Delta P_{qq} (x) & = & C_F \left[\frac{2}{(1-x)_{+}} -1-x +\frac{3}{2}\delta 
(1-x)\right], \qquad \Delta z_{qq}(x)  =-4C_F (1-x).\\
\Delta P_{qg} (x) & =&  T_F \left[ 2x-1 \right].
\end{eqnarray*}

After mass factorization, the 
polarized cross sections for the Drell--Yan production of lepton pairs 
at NLO can be expressed in a form similar to the unpolarized cross 
sections~\cite{smrs}:
\begin{eqnarray}
\frac{\d \Delta \sigma}{\d M^2 \d x_F} & = & \frac{4\pi \alpha^2}{9 M^2 S} 
\sum_i e_i^2 \int_{x_1^0}^1 \d x_1 \int_{x_2^0}^1 \d x_2 \nonumber \\
& & \hspace{-0.7cm}
\times \Bigg\{\left[\frac{\d \Delta \hat{\sigma}_{q\bar{q}}^{(0)}}
{\d M^2 \d x_F} (x_1,x_2) +\frac{\alpha_s}{2\pi}
\frac{\d \Delta \hat{\sigma}_{q\bar{q}}^{(1)}}
{\d M^2 \d x_F}\left(x_1,x_2,\frac{M^2}{\mu_F^2}\right)\right]
\Big\{ \Delta q_i(x_1,\mu_F^2)\Delta\bar{q}_i(x_2,\mu_F^2) 
\nonumber \\
& & \hspace{0.1cm}
+\Delta\bar{q}_i(x_1,\mu_F^2)\Delta q_i(x_2,\mu_F^2) \Big\} \nonumber \\
& & \hspace{-0.2cm} + \Bigg[ \frac{\alpha_s}{2\pi}
\frac{\d \Delta \hat{\sigma}_{qg}^{(1)}} {\d M^2 \d x_F} \left(x_1,x_2,
\frac{M^2}{\mu_F^2}\right) \Delta G(x_1,\mu_F^2) \left\{ 
\Delta q_i(x_2,\mu_F^2) + 
\Delta \bar{q}_i (x_2,\mu_F^2) \right\} \nonumber \\
& & \hspace{0.3cm}
+ (1 \leftrightarrow 2)\Bigg] \Bigg\},
\label{eq:xfmaster}
\end{eqnarray}
with 
\begin{eqnarray}
\frac{\d \Delta \hat{\sigma}_{q\bar{q}}^{(0)}}
{\d M^2 \d x_F} (x_1,x_2) & = & -\frac{\delta (x_1-x_1^0)\, 
\delta (x_2-x_2^0)}{x_1^0+x_2^0} = - \frac{\d \hat{\sigma}_{q\bar{q}}^{(0)}}
{\d M^2 \d x_F} (x_1,x_2)\; , \\
\frac{\d \Delta \hat{\sigma}_{q\bar{q}}^{(1)}}
{\d M^2 \d x_F} \left(x_1,x_2,\frac{M^2}{\mu_F^2}\right)
& = & -C_F \Bigg\{ \frac{\delta (x_1-x_1^0) \,
\delta (x_2-x_2^0)}{x_1^0+x_2^0} \bigg[ \frac{\pi^2}{3} - 8 + 2 \Li_2 (\xaa ) 
+ 2 \Li_2 ( \xbb ) 
\nonumber \\ 
& & \hspace{0.6cm}
+\ln^2 (1-\xaa) + \ln^2 (1-\xbb) + 2 \ln \frac{\xaa}{1-\xaa} \ln 
\frac{\xbb}{1-\xbb} \bigg]  \nonumber \\
& & + \Bigg(\frac{\delta (\xa-\xaa)}{x_1^0+x_2^0} \bigg[ \frac{1}{\xb} 
- \frac{\xbb}{\xb^2} - \frac{\xbb\,^2+\xb^2}{\xb^2(\xb-\xbb)} \ln 
\frac{\xbb}{\xb} \nonumber \\
& & \hspace{0.6cm}
+ \frac{\xbb\,^2+\xb^2}{\xb^2} 
\left(\frac{\ln (1-\xbb/\xb)}{\xb-\xbb}\right)_{+}
+ \frac{\xbb\,^2+\xb^2}{\xb^2} \frac{1}{\left(\xb-\xbb\right)_{+}}
\nonumber\\ && \hspace{0.6cm} \ln \frac
{(\xaa+\xbb)(1-\xaa)}{\xaa (\xaa+\xb)} 
\bigg] + (1 \leftrightarrow 2 ) \Bigg) \nonumber \\
& & + \frac{\Delta
\tilde{G}^A(\xa,\xb)}{\left[(\xa-\xaa)(\xb-\xbb)\right]_{+}} + 
\Delta\tilde{H}^A (\xa,\xb) 
\nonumber \\
&& 
+ \ln \frac{M^2}{\mu_F^2} \Bigg\{  \frac{\delta (x_1-x_1^0)\, 
\delta (x_2-x_2^0)}{x_1^0+x_2^0} \bigg[ 3 + 2 \ln \frac{1-\xaa}{\xaa} + 2 \ln 
\frac{1-\xbb}{\xbb}
\bigg] \nonumber\\
&& \hspace{0.6cm}
+ \bigg( \frac{\delta (\xa-\xaa)}{x_1^0+x_2^0}   
 \frac{\xbb\,^2+\xb^2}{\xb^2}\frac{1}{\left(\xb-\xbb\right)_{+}} + (1 
\leftrightarrow 2 ) \bigg) \Bigg\} \Bigg\}   \\
&=&- \frac{\d \hat{\sigma}_{q\bar{q}}^{(1)}}
{\d M^2 \d x_F} \left(x_1,x_2,\frac{M^2}{\mu_F^2}\right)\; , \nonumber\\
\frac{\d \Delta \hat{\sigma}_{qg}^{(1)}}
{\d M^2 \d x_F} \left(x_1,x_2,\frac{M^2}{\mu_F^2}\right) & = & 
-T_F\Bigg\{\frac{\delta (\xb-\xbb) }{(\xaa+\xbb)\xa^2} \Bigg[ (2\xaa -\xa) 
\ln\frac{(\xaa+\xbb)(1-\xbb)(\xa-\xaa)}{\xaa\xbb(\xa+\xbb)} 
 \nonumber \\
& & \hspace{0.6cm}
+ 2(\xa-\xaa)\Bigg] 
+ \frac{\Delta\tilde{G}^C(\xa,\xb)}{(\xb-\xbb)_{+}} + 
\Delta\tilde{H}^C(\xa,\xb) 
\nonumber \\ & & 
+  \ln \frac{M^2}{\mu_F^2} \Bigg\{ \frac{\delta (\xb-\xbb) }{(\xaa+\xbb)\xa^2}
 (2\xaa -\xa) \Bigg\}\Bigg\}\; , 
\end{eqnarray}
where
\begin{eqnarray}
\Delta\tilde{G}^A(\xa,\xb) & = & \frac{(\xa+\xb)(\xaa\,^2\xbb\,^2+\xa^2\xb^2)}
{\xa^2\xb^2(\xaa+\xb)(\xa+\xbb)}\; , \nonumber \\
\Delta\tilde{H}^A (\xa,\xb) & = & 
-\frac{2}{\xa\xb(\xa+\xb)}\; , \nonumber \\
\Delta\tilde{G}^C(\xa,\xb) & = & \frac{2\xaa\xbb-\xa\xb}{\xa^2\xb(\xaa+\xb)}
\; ,\nonumber \\
\Delta\tilde{H}^C (\xa,\xb) & = &\frac{\xa(\xaa+\xb)(\xb-\xbb)+2\xaa\xbb
(\xa+\xb)}{\xa^2\xb^2(\xa+\xb)^2} \; . \nonumber  
\end{eqnarray}

Inserting 
\begin{displaymath}
\int \d y \,\delta \left(y - \frac{1}{2}\left[ \ln \frac{\xa}{\xb} + \ln 
\frac{z+z_1-zz_1}{1-z_1+zz_1} \right]\right) 
\end{displaymath}
into eqs.(\ref{eq:qqbare})--(\ref{eq:qgbare})
and using identities (\ref{eq:delx3}),(\ref{eq:dely1})--(\ref{eq:dely3}),
one can derive the ${\cal O}(\alpha_s)$--corrections to the rapidity 
distribution of the Drell--Yan pairs. The variables 
\begin{equation}
\xaa = \sqrt{\tau} e^{y}, \qquad \xbb = \sqrt{\tau} e^{-y}
\label{eq:ydef}
\end{equation}
are in this case defined by 
\begin{displaymath}
y = \frac{1}{2} \ln \frac{\xaa}{\xbb} \qquad \mbox{and} \qquad \tau =\xaa\xbb.
\end{displaymath}
They are in general different from the $\xaa$ and $\xbb$ occuring in the
$x_F$-distributions. 

These corrections have already
been calculated in \cite{weber}; we find  numerical agreement with the 
results (keeping in mind that a different factorization prescription for
the gluonic contribution was used in \cite{weber}), our expressions are 
however considerably simpler:
\begin{eqnarray}
\frac{\d \Delta \sigma}{\d M^2 \d y} & = & \frac{4\pi \alpha^2}{9 M^2 S} 
\sum_i e_i^2 \int_{x_1^0}^1 \d x_1 \int_{x_2^0}^1 \d x_2 \nonumber \\
& & \hspace{-0.7cm}
\times \Bigg\{\left[\frac{\d \Delta \hat{\sigma}_{q\bar{q}}^{(0)}}
{\d M^2 \d y} (x_1,x_2) +\frac{\alpha_s}{2\pi}
\frac{\d \Delta \hat{\sigma}_{q\bar{q}}^{(1)}}
{\d M^2 \d y}\left(x_1,x_2,\frac{M^2}{\mu_F^2}\right)\right]
\Big\{ \Delta q_i(x_1,\mu_F^2)\Delta\bar{q}_i(x_2,\mu_F^2) 
\nonumber \\
& & \hspace{0.1cm}
+\Delta\bar{q}_i(x_1,\mu_F^2)\Delta q_i(x_2,\mu_F^2) \Big\} \nonumber \\
& & \hspace{-0.2cm} + \Bigg[ \frac{\alpha_s}{2\pi}
\frac{\d \Delta \hat{\sigma}_{qg}^{(1)}} {\d M^2 \d y} \left(x_1,x_2,
\frac{M^2}{\mu_F^2}\right) \Delta G(x_1,\mu_F^2) \left\{ 
\Delta q_i(x_2,\mu_F^2) + 
\Delta \bar{q}_i (x_2,\mu_F^2) \right\} \nonumber \\
& & \hspace{0.3cm}
+ (1 \leftrightarrow 2)\Bigg] \Bigg\},
\label{eq:ymaster}
\end{eqnarray}
with 
\begin{eqnarray}
\frac{\d \Delta \hat{\sigma}_{q\bar{q}}^{(0)}}
{\d M^2 \d y} (x_1,x_2) & = & -\delta (x_1-x_1^0)\, 
\delta (x_2-x_2^0) = - \frac{\d \hat{\sigma}_{q\bar{q}}^{(0)}}
{\d M^2 \d y} (x_1,x_2) \; , \\
\frac{\d \Delta \hat{\sigma}_{q\bar{q}}^{(1)}}
{\d M^2 \d y} \left(x_1,x_2,\frac{M^2}{\mu_F^2}\right)
& = & -C_F \Bigg\{ \delta (x_1-x_1^0) \,
\delta (x_2-x_2^0) \bigg[ \frac{\pi^2}{3} - 8 + 2 \Li_2 (\xaa ) 
+ 2 \Li_2 ( \xbb ) 
\nonumber \\ 
& & \hspace{0.6cm}
+\ln^2 (1-\xaa) + \ln^2 (1-\xbb) + 2 \ln \frac{\xaa}{1-\xaa} \ln 
\frac{\xbb}{1-\xbb} \bigg]  \nonumber \\
& & + \Bigg( \delta (\xa-\xaa) \bigg[ \frac{1}{\xb} 
- \frac{\xbb}{\xb^2} - \frac{\xbb\,^2+\xb^2}{\xb^2(\xb-\xbb)} \ln 
\frac{\xbb}{\xb} \nonumber \\
& & \hspace{0.6cm}
+ \frac{\xbb\,^2+\xb^2}{\xb^2} 
\left(\frac{\ln (1-\xbb/\xb)}{\xb-\xbb}\right)_{+}
+ \frac{\xbb\,^2+\xb^2}{\xb^2} \frac{1}{\left(\xb-\xbb\right)_{+}}
\nonumber\\ && \hspace{0.6cm} \ln \frac
{2\xbb(1-\xaa)}{\xaa(\xb+\xbb)} 
\bigg] + (1 \leftrightarrow 2 ) \Bigg) \nonumber \\
& & + \frac{\Delta
G^A(\xa,\xb)}{\left[(\xa-\xaa)(\xb-\xbb)\right]_{+}} + 
\Delta H^A (\xa,\xb) 
\nonumber \\
&& 
+ \ln \frac{M^2}{\mu_F^2} \Bigg\{  \delta (x_1-x_1^0)\, 
\delta (x_2-x_2^0) \bigg[ 3 + 2 \ln \frac{1-\xaa}{\xaa} + 2 \ln 
\frac{1-\xbb}{\xbb}
\bigg] \nonumber\\
&& \hspace{0.6cm}
+ \bigg( \delta (\xa-\xaa)  
 \frac{\xbb\,^2+\xb^2}{\xb^2}\frac{1}{\left(\xb-\xbb\right)_{+}} + (1 
\leftrightarrow 2 ) \bigg) \Bigg\} \Bigg\}   \\
&=&- \frac{\d \hat{\sigma}_{q\bar{q}}^{(1)}}
{\d M^2 \d y} \left(x_1,x_2,\frac{M^2}{\mu_F^2}\right)\; , \nonumber \\
\frac{\d \Delta \hat{\sigma}_{qg}^{(1)}}
{\d M^2 \d y} \left(x_1,x_2,\frac{M^2}{\mu_F^2}\right) & = & 
-T_F\Bigg\{\frac{\delta (\xb-\xbb) }{\xa^2} \Bigg[ (2\xaa -\xa) 
\ln\frac{2(\xa-\xaa)(1-\xbb)}{(\xa+\xaa)\xbb} 
 \nonumber \\
& & \hspace{0.6cm}
+ 2(\xa-\xaa)\Bigg] 
+ \frac{\Delta G^C(\xa,\xb)}{(\xb-\xbb)_{+}} + 
\Delta H^C(\xa,\xb) 
\nonumber \\ & & 
+  \ln \frac{M^2}{\mu_F^2} \Bigg\{ \frac{\delta (\xb-\xbb) }{\xa^2}
 (2\xaa -\xa) \Bigg\}\Bigg\}\; , 
\end{eqnarray}
where
\begin{eqnarray}
\Delta G^A(\xa,\xb) & = & \frac{2 (\xa\xb+\xaa\xbb)
(\xaa\,^2\xbb\,^2+\xa^2\xb^2)}
{\xa^2\xb^2(\xa+\xaa)(\xb+\xbb)}\; , \nonumber \\
\Delta H^A (\xa,\xb) & = & 
-\frac{4\xaa\xbb(\xaa\xbb+\xa\xb)}{\xa\xb(\xa\xbb+\xb\xaa)^2}\; , \nonumber \\
\Delta G^C(\xa,\xb) & = & \frac{2\xbb(2\xaa\xbb-\xa\xb)(\xaa\xbb+\xa\xb)}
{\xa^2\xb(\xa\xbb+\xb\xaa)(\xb+\xbb)}\; ,
\nonumber \\
\Delta H^C (\xa,\xb) & = &\frac{2\xaa\xbb(\xaa\xbb+\xa\xb)(\xa\xaa\xb^2+
\xaa\xbb(\xa\xbb+2\xaa\xb))}{\xa^2\xb^2(\xa\xbb+\xb\xaa)^3}\; . \nonumber  
\end{eqnarray}  
As mentioned above, the ${\cal O}(\alpha_s)$--corrections to the $x_F$- 
and $y$-distributions in the unpolarized Drell-Yan process
were derived in~\cite{aem,kubar}. A summary of these 
formulae in the $\overline{{\rm MS}}$--scheme in our notation can be found 
in~\cite{smrs}\footnote{As the normalization of the gluon polarization sum
used in~\cite{smrs,aem,kubar} 
deviates from the normalization conventionally used in
dimensional regularization, eqs.(A8) and (A20) of~\cite{smrs}
have to be modified according to~\cite{grvdy} to conform to
the normalization used in the NLO splitting functions and 
parton distributions.}. 

Finally, integration of eqs.(\ref{eq:qqbare})--(\ref{eq:qgbare})
with respect to $z_1$ reproduces after mass factorization using 
eq.~(\ref{eq:mfac}) 
the correction terms to the Drell--Yan mass distribution~\cite{ratcliffe,weber,
kamal}. These read in the $\overline{{\rm MS}}$--scheme:
\begin{eqnarray}
\lefteqn{\frac{\d \Delta  \sigma}{\d M^2}  =  \frac{4 \pi \alpha^2}{9 S M^2}
\int_{0}^{1} \d x_1 \d x_2 \d z \delta (x_1x_2z-\tau) 
\sum_q e_q^2 } \nonumber \\
& & \hspace{-0.5cm} \times \Bigg\{ 
\left\{\, \Delta q(x_1,\mu_F^2)\, \Delta
\barq (x_2,\mu_F^2) + (1 \leftrightarrow 2) \, \right\} \left( 
\Delta c_{q\barq}^{DY,(0)} (z)   + \frac{\alpha_s}{2 \pi}  
\Delta c_{q\barq}^{DY,(1)} \left(z,\frac{M^2}{\mu_F^2}\right)
 \right) \nonumber \\
&& \hspace{-0.5cm} + \left\{\left( \, \Delta 
q(x_1,\mu_F^2)+ \Delta \bar{q}(x_1,\mu_F^2)\right)\, \Delta
G(x_2,\mu_F^2) + (1
\leftrightarrow 2) \, \right\}   \frac{\alpha_s}{2 \pi}
\Delta c_{qg}^{DY,(1)} \left(z,\frac{M^2}{\mu_F^2}\right)
\Bigg\},  \label{eq:signloDY}
\end{eqnarray}
with:
\begin{eqnarray}
\Delta c_{q\barq}^{DY,(0)} (z) & = & -\delta(1-z) = -  c_{q\barq}^{DY,(0)} (z)
 \nonumber \\
\Delta c_{q\barq}^{DY,(1)} \left(z,\frac{M^2}{\mu_F^2}\right) 
& = & -C_F   \Bigg[  8\,  \left(\frac{\ln
(1-z)}{1-z}\right)_{+} 
-2\,\frac{1+z^2}{1-z} \,\ln z - 4 (1+z) \ln (1-z)
\nonumber \\
& & \hspace{1.2cm} + \delta (1-z) \left(-8 + \frac{2\pi^2}{3} \right)
\nonumber \\
& & \hspace{1.2cm}
+ 2 \ln \frac{M^2}{\mu_F^2} \left\{
\left(\frac{2}{1-z}\right)_{+} -1-z+\frac{3}{2} \delta (1-z) \right\}
\Bigg] \, , \nonumber \\ 
& = & - c_{q\barq}^{DY,(1)} \left(z,\frac{M^2}{\mu_F^2}\right)\nonumber \\ 
\Delta c_{qg}^{DY,(1)} \left(z,\frac{M^2}{\mu_F^2}\right) & = & -T_F \Bigg[
\left(2z-1\right) \ln \frac{(1-z)^2}{z}
+\frac{5}{2}-z-\frac{3}{2} z^2 + \ln
\frac{M^2}{\mu_F^2} \left\{2z-1\right\} \Bigg] \, .\nonumber
\end{eqnarray} 
 
\section{Numerical results}
\label{sec:num}
In the following, we will illustrate the impact of the QCD 
corrections to the polarized Drell--Yan process derived above by numerically
evaluating (\ref{eq:xfmaster}) and (\ref{eq:ymaster}) for recent 
parameterizations of the polarized parton distribution functions.

When using the polarized GS parton distribution 
functions~\cite{gs},
we take $\Lambda^{{\rm QCD}}_{n_f=4}=231$~MeV, the 
corresponding unpolarized cross sections 
are then evaluated using the unpolarized MRS parton
distribution functions set A$'$~\cite{mrsap}. The polarized 
GRSV distributions~\cite{grsv} are 
consistently used in combination with the unpolarized distributions 
from GRV~\cite{grv} and for
$\Lambda^{{\rm QCD}}_{n_f=4}=200$~MeV. If not stated otherwise, we shall 
always use $\mu_F=M$; the strong coupling constant $\alpha_s$ is evaluated 
at $\mu_F$. 

All results in this section are obtained for $\sqrt{s}=39.22$~GeV, 
corresponding to a fixed target experiment in the HERA proton beam~\cite{HERA}
with $E_{beam}=820$~GeV.

\begin{figure}[t]
\begin{center}
~ \epsfig{file=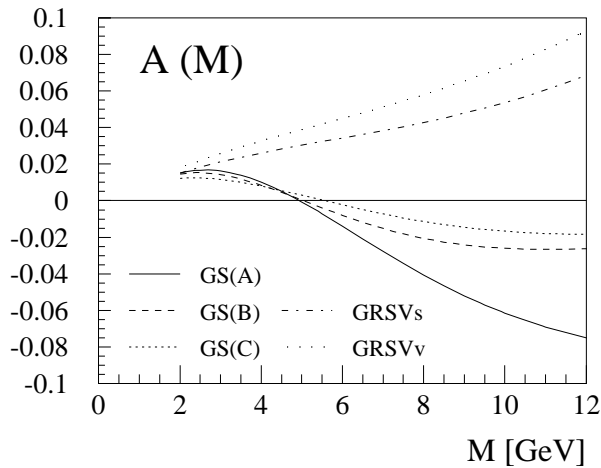,angle=-90,width=8cm}
\caption{The Drell--Yan asymmetry $A(M)$ in $pp$ collisions with 
$E_{beam}=820$ GeV for different parameterizations of the polarized parton 
distribution functions.} 
\label{fig:fig2}
\end{center}
\end{figure}

At present, all polarized parton distributions are fitted to data on the 
polarized structure function $g_1(x,Q^2)$ only. As this structure function is
dominated by valence quark contributions for $x>0.1$, the knowledge on the 
polarized sea quark distributions in the large-$x$ region is very poor.
This uncertainty is illustrated in Fig.~\ref{fig:fig2}, which shows the
next-to-leading order predictions (Eq.~(\ref{eq:signloDY}))
for the Drell--Yan asymmetry 
\begin{displaymath}
A(M) \equiv \frac{\d \Delta \sigma}{\d M} \Bigg/
\frac{\d \sigma}{\d M}
\end{displaymath}
obtained using the polarized parton distribution functions of~\cite{gs} 
(GS(A--C)) and~\cite{grsv} (GRSVs,v). At present, neither magnitude nor
sign of this asymmetry can be predicted. It should therefore be clear,
that all numerical results for the polarized Drell--Yan process
presented in the remainder of this section do not make solid predictions for 
observable asymmetries. The purpose of these numerical calculations is only to
illustrate the magnitude of the next-to-leading order corrections and to 
demonstrate the perturbative stability of the predictions.

\begin{figure}[t]
\begin{center}
~ \epsfig{file=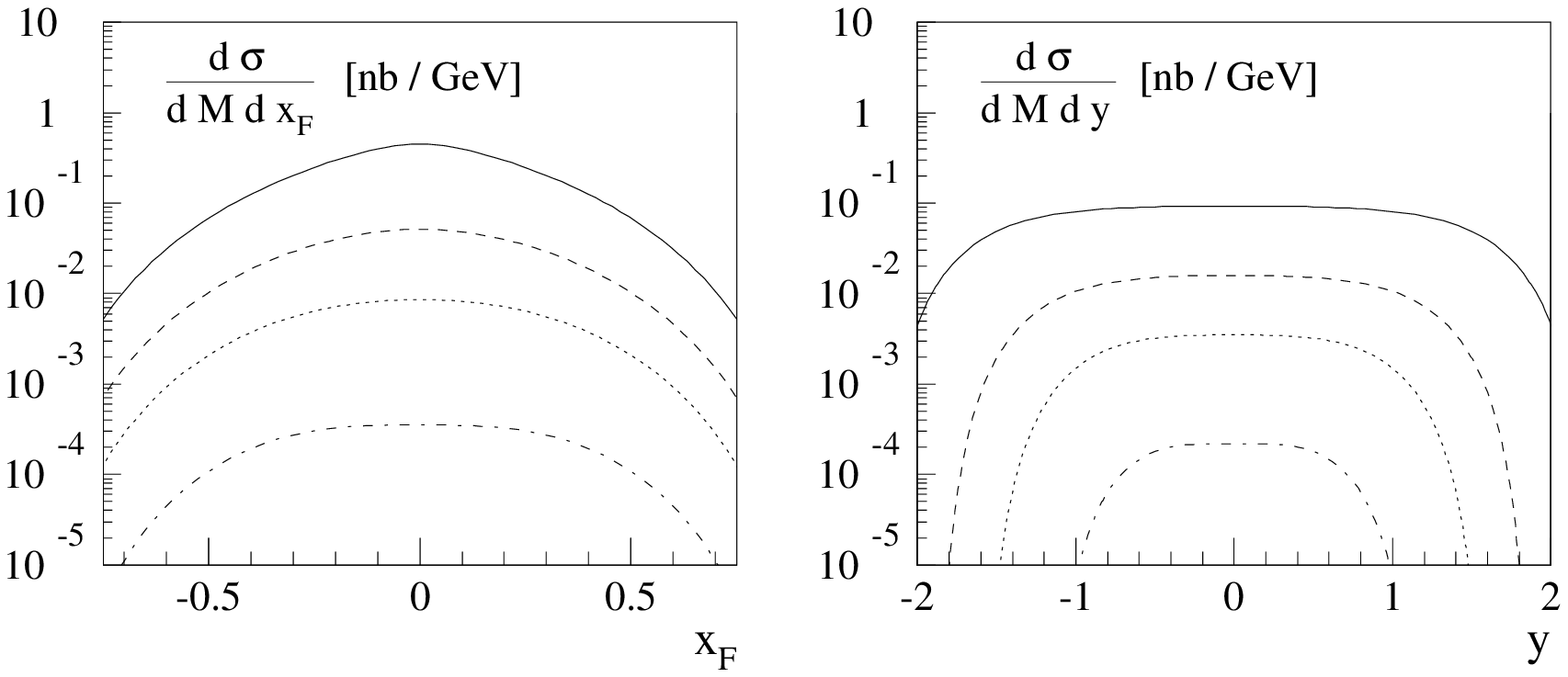,angle=-90,width=16cm}
\caption{The unpolarized Drell--Yan cross section in $pp$ collisions. Curves 
are for different invariant masses $M$; solid: 4~GeV, dashed: 6~GeV, dotted:
8~GeV, dot-dashed: 12~GeV.} 
\label{fig:fig3}
~ \epsfig{file=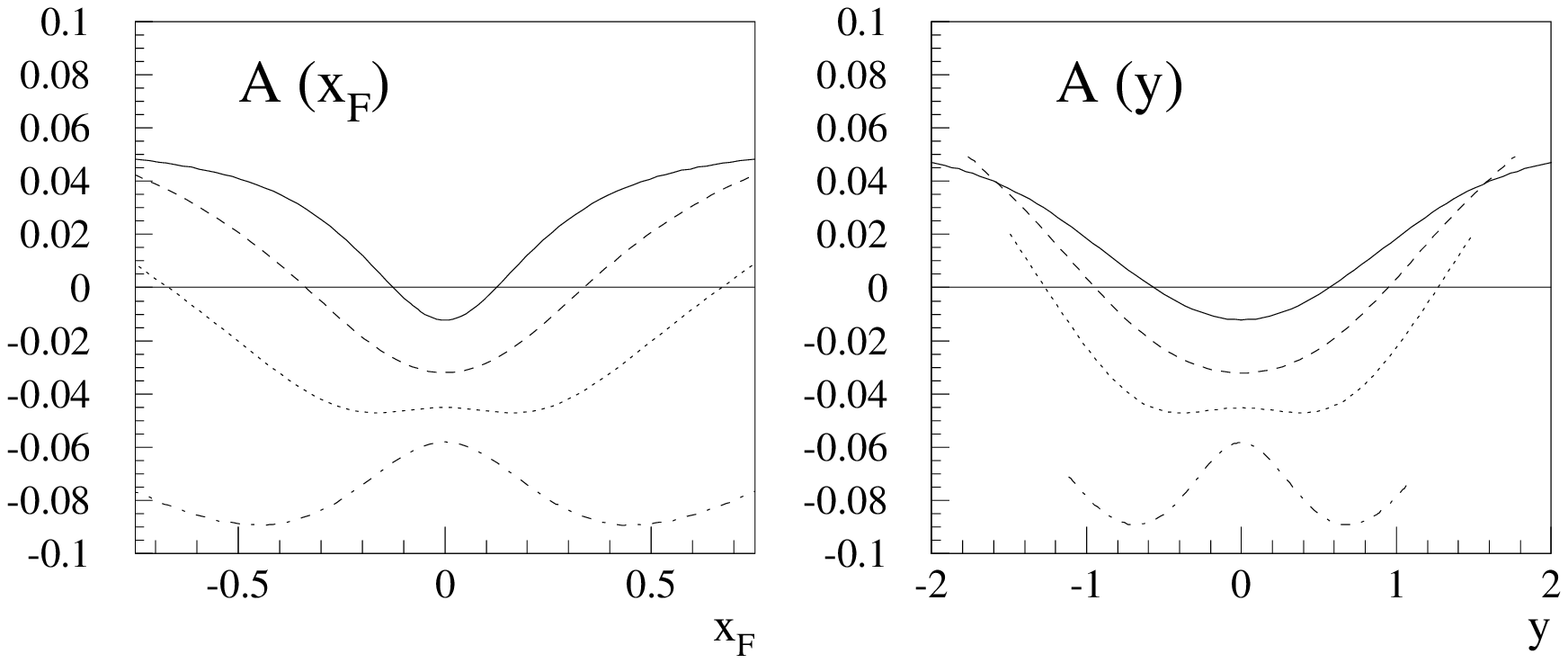,angle=-90,width=16cm}
\caption{The Drell--Yan asymmetry in $pp$ collisions for the polarized 
GS(A) parton distributions. Curves 
are for different invariant masses $M$; solid: 4~GeV, dashed: 6~GeV, dotted:
8~GeV, dot-dashed: 12~GeV.} 
\label{fig:fig4}
\end{center}
\end{figure}

In the following, we shall always study the cross section asymmetries
\begin{displaymath}
A(x_F) \equiv \frac{\d \Delta \sigma}{\d M \d x_F}\, \Bigg/ \,
\frac{\d \sigma}{\d M \d x_F}
\qquad \mbox{and}
\qquad A(y) \equiv \frac{\d \Delta \sigma}{\d M \d y}\, \Bigg/ \,
\frac{\d \sigma}{\d M \d y}\;.
\end{displaymath}
As the unpolarized parton distributions are known to some level of accuracy,
it is possible to predict the cross sections in the denominator of $A(x_F)$
and $A(y)$.
These cross sections are shown for proton--proton 
collisions ($\sqrt{s}=39.22$~GeV) in Fig.~\ref{fig:fig3}, which is 
obtained with the MRS(A$'$) parton distributions~\cite{mrsap}. The prediction
using the GRV parton distributions~\cite{grv} agrees within 3\% over the 
whole kinematic range.

It can be seen from Fig.~\ref{fig:fig2} that the total Drell--Yan asymmetry
$A(M)$ changes with the invariant mass $M$. 
The change is however not uniform over the whole range in $x_F$ or $y$ as 
illustrated in Fig.~\ref{fig:fig4} for the polarized GS(A) parton 
distributions. The $x_F$-dependence of the asymmetry 
can be understood from eq.~(\ref{eq:xfdef}).
The asymmetry around $x_F=0$ always probes the 
polarized parton distributions in beam~$(1)$ and target~$(2)$ 
around $\xaa\approx \xbb
\approx \sqrt{\tau}$. If $x_F$ is increased towards its positive
kinematic limit, $\xaa$ increases towards 1 while $\xbb$ decreases only
slightly. Decreasing $x_F$ towards the negative kinematic limit yields a 
small decrease of $\xaa$ while $\xbb$ increases towards 1. The region 
$x_F>0$ therefore probes the $x$-dependence 
of the parton distributions in the beam hadron, while $x_F<0$ is sensitive
on the target hadron structure. Moreover, different invariant masses 
correspond to different $x$ intervals probed. 
A similar argumentation, based on 
eq.~(\ref{eq:ydef}), applies to the $y$-dependence of the asymmetry. 
\begin{figure}[t]
\begin{center}
~ \epsfig{file=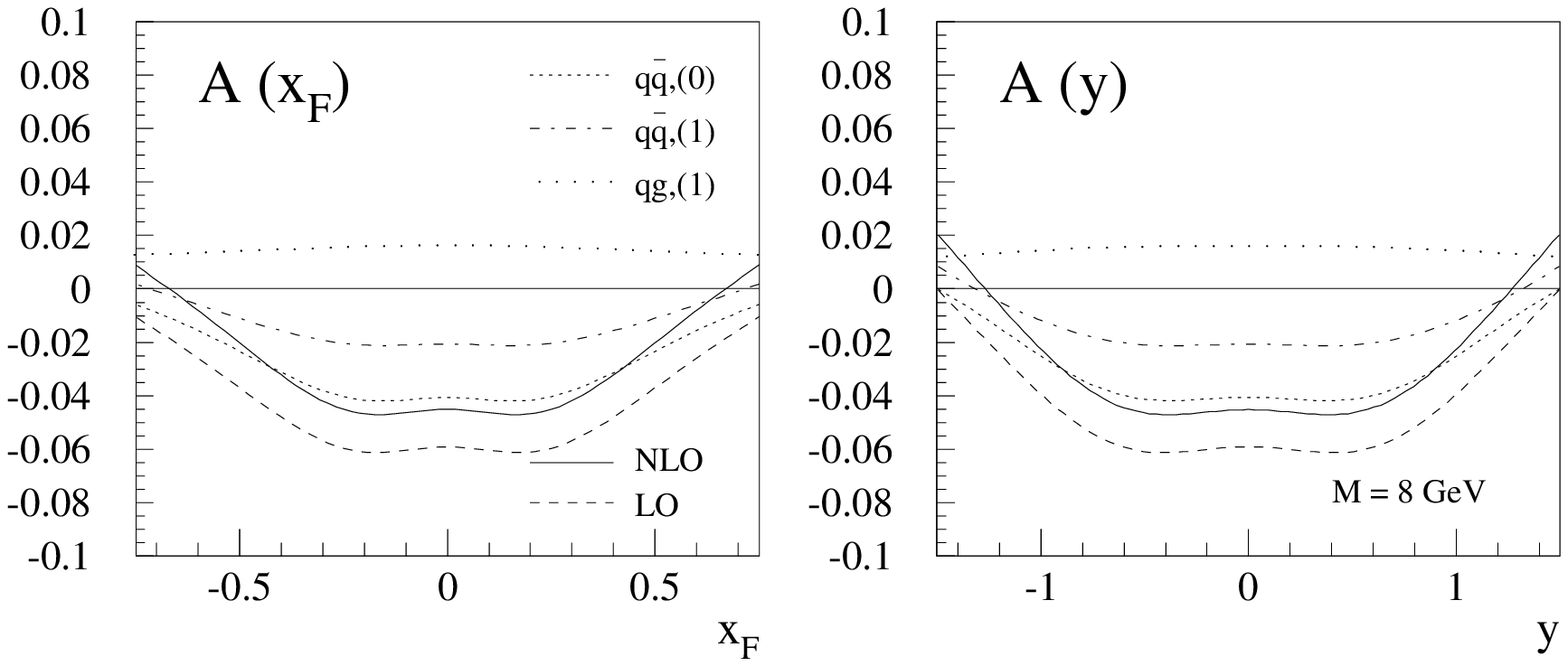,angle=-90,width=16cm}
\caption{Contributions of the individual parton level subprocesses to the 
polarized Drell--Yan cross section. The subprocess cross sections 
(short-dashed, dot-dashed and dotted lines) and their sum (solid line) are
normalized to the full unpolarized cross section at NLO. The long-dashed 
line shows the asymmetry evaluated at leading order.} 
\label{fig:fig5}
~ \epsfig{file=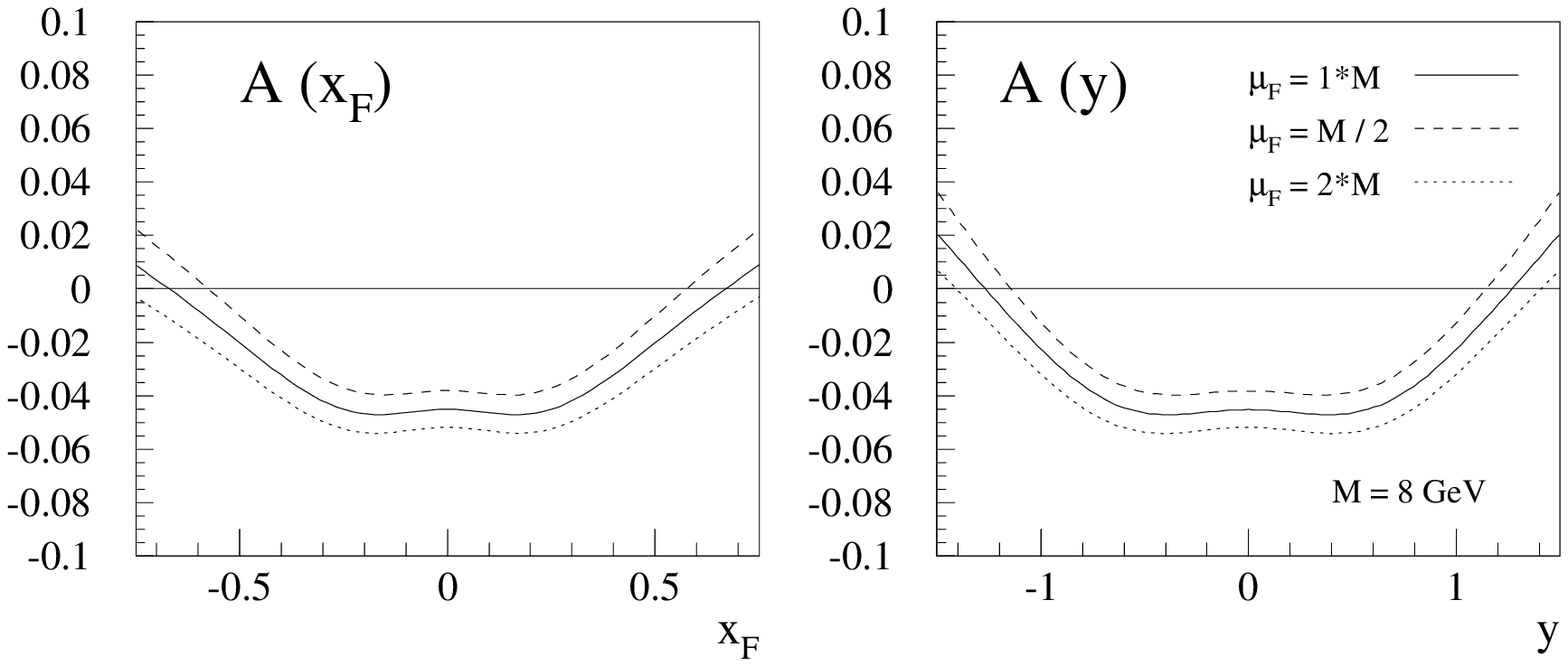,angle=-90,width=16cm}
\caption{Sensitivity of the Drell--Yan asymmetry on variations of the 
factorization scale.} 
\label{fig:fig6}
\end{center}
\end{figure}

The contributions of the individual subprocesses ($q\barq$--annihilation at
leading and next-to-leading order and quark--gluon Compton scattering) to 
the polarized Drell--Yan cross section are illustrated in Fig.~\ref{fig:fig5}.
All curves are obtained with the polarized GS(A) parton densities and are 
shown for $M=8$~GeV.
All polarized 
subprocess contributions are normalized to the {\it full} unpolarized 
cross section at next-to-leading order. The relative magnitude of the 
individual contributions is similar to the unpolarized Drell--Yan process.
The ${\cal O}(\alpha_s)$ correction to the $q\barq$--annihilation process 
enhances significantly the lowest order prediction while the quark--gluon 
Compton process contributes with a sign opposite to the annihilation process. 
However, the relative magnitude of annihilation and Compton process depends 
on the magnitude of the gluon distribution at large $x$, which is 
completely undetermined at present~\cite{gs,grsv}. This uncertainty 
prevents a sensible prediction of a
$K$--factor between the cross sections at leading and
next-to-leading order. 

Furthermore, Fig.~\ref{fig:fig5} shows the prediction for the Drell--Yan 
asymmetry if both polarized and unpolarized cross section are evaluated 
using the leading order $q\barq,(0)$-coefficient function only
(long--dashed line). 
The sizable shift between the leading and next-to-leading order predictions 
for the Drell--Yan asymmetry seems to contradict the claim~\cite{ratcliffe}
that the next-to-leading order corrections, although large in polarized 
and unpolarized cross sections, would largely cancel in the Drell--Yan 
asymmetry.

The change of physical 
quantities such as cross sections or asymmetries under variations of the 
(unphysical) mass factorization scale $\mu_F$ enables an estimate of the 
numerical importance of unknown higher order corrections. We quantify
this uncertainty in Fig.~\ref{fig:fig6}, showing the Drell--Yan asymmetry
for $\mu_F=0.5,\,1,\,2\, M$. It can be seen that
the absolute value of the asymmetry changes less
than 0.01 in the central region.
Even in the region where $x_F$ (or $y$) approaches its kinematic limit, the 
absolute value of the asymmetry changes only by a maximum of 0.015.  
This stability is crucial for an estimate of the theoretical error
on the polarized sea quark distributions extracted from the polarized 
Drell--Yan process. 
\begin{figure}[t]
\begin{center}
~ \epsfig{file=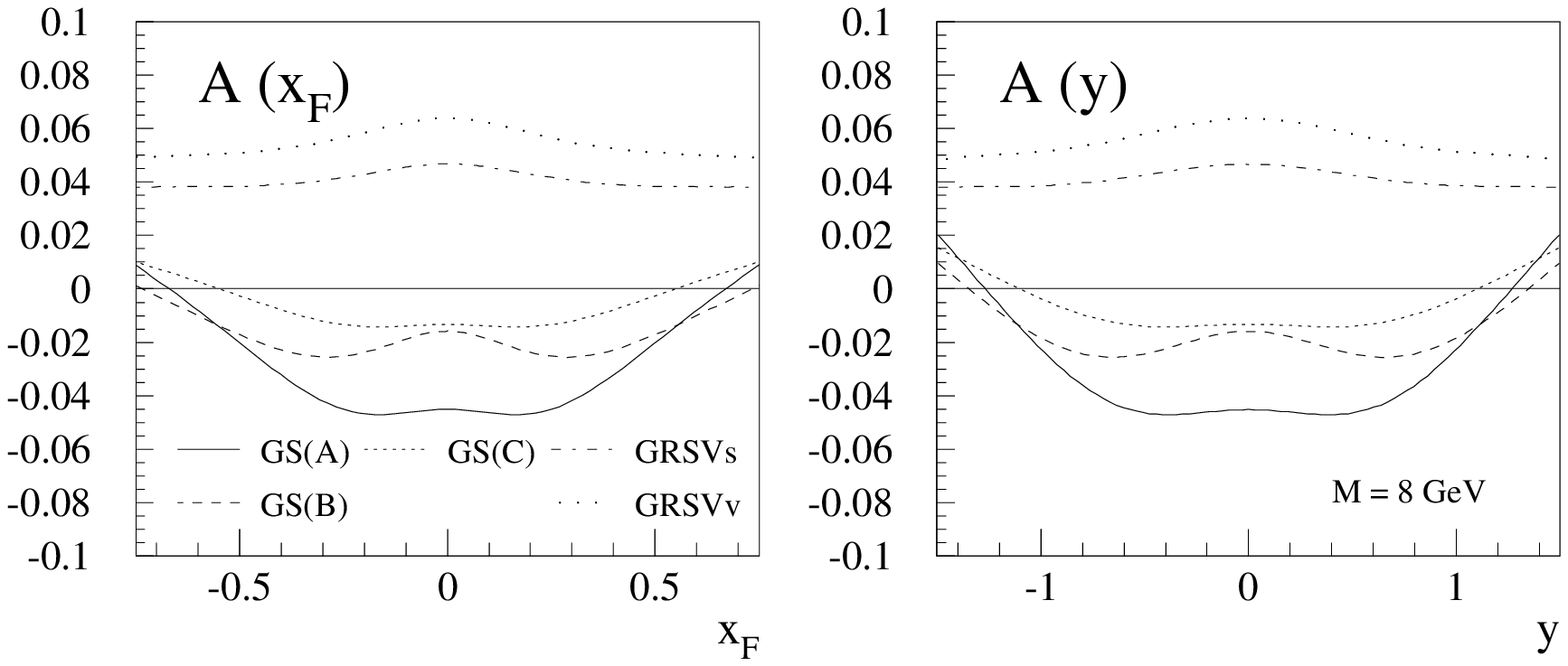,angle=-90,width=16cm}
\caption{Predictions for the Drell--Yan asymmetry at $M=8$~GeV 
in polarized $pp$ collisions 
($E_{beam}=820$~GeV) obtained for different parameterizations of 
the polarized parton densities. } 
\label{fig:fig7}
~ \epsfig{file=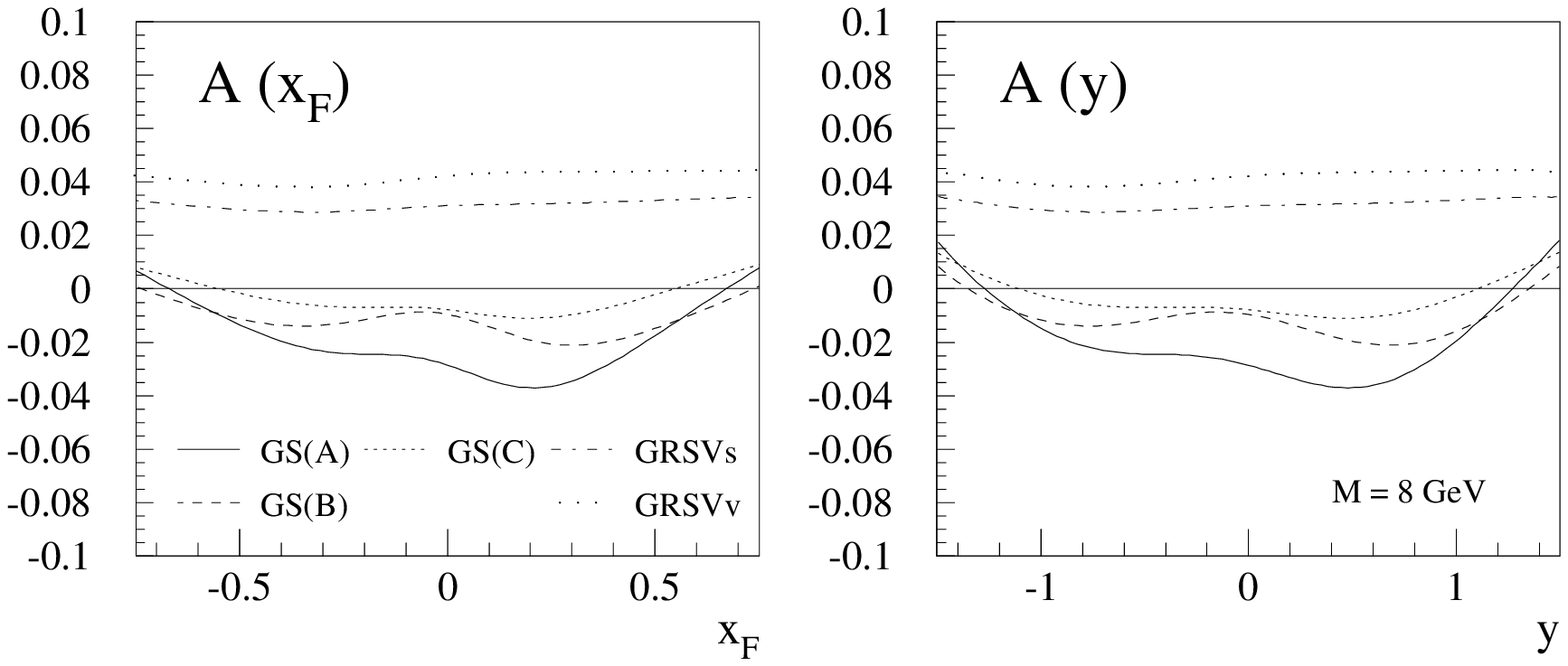,angle=-90,width=16cm}
\caption{Predictions for the Drell--Yan asymmetry at $M=8$~GeV 
in polarized $pd$ collisions 
($E_{beam}=820$~GeV) obtained for different parameterizations 
of the polarized parton densities. The proton is taken in $x_1$--direction.} 
\label{fig:fig8}
\end{center}
\end{figure}

A final point briefly concerns phenomenological aspects of the polarized 
Drell--Yan process, although this is not the principal 
aim of the present paper.  Figures~\ref{fig:fig7} and~\ref{fig:fig8}
show the Drell--Yan asymmetry in polarized 
proton--proton and proton--deuterium
collisions for $M=8$~GeV as predicted by using the different 
polarized parton distributions suggested in~\cite{gs} and~\cite{grsv}. 
It can clearly be seen that the different predictions vary within a band
of $\pm 0.06$ in $pp$ and $\pm 0.04$ in $pd$ collisions.
This uncertainty reflects the present lack of knowledge on the 
polarization of the light quark sea in the proton at $x\gapprox 0.1$. The 
difference between the predictions is furthermore substantially larger 
than the theoretical uncertainty due to the choice of 
the factorization scale. 

Due to the smaller overall valence quark polarization in deuterium,
one finds the asymmetry in $pd$ collisions to be 
 systematically smaller than in $pp$ collisions. 
Although the normalization of the asymmetry in the proton direction
($x_F,y>0$)
and in the deuteron direction ($x_F,y<0$) is slightly different, its shape 
is still very similar in both regions. This reflects the assumption of an 
SU(2)$_f$--symmetry of the polarized antiquark sea, which is used in all
polarized parton distributions displayed in the figure. A measurement of 
the $x_F$-(or $y$-)dependence of the Drell--Yan asymmetry in $pd$ collisions
could therefore test this assumption. 
\goodbreak

\section{Conclusions}
\label{sec:conc}
We have presented a complete calculation of the ${\cal O}(\alpha_s)$ 
corrections to the $x_F$- and $y$-dependence of the longitudinally 
polarized Drell--Yan cross section in the $\overline{{\rm MS}}$--scheme.
The results for the $y$-dependence and
the mass distribution agree (after suitable change of the factorization 
scheme) with earlier results in the literature~\cite{ratcliffe,weber,kamal}. 
We have demonstrated that the impact of these corrections on the polarized 
Drell--Yan cross section is very similar to the impact of the unpolarized 
corrections. The 
corrections due to the individual subprocesses
turn out to be sizable. However, a $K$--factor
between leading and next-to-leading order results can at present
not be sensibly predicted due to the uncertainty on the magnitude of the 
polarized gluon distribution. 

The corrections derived in this paper are of particular importance for 
the extraction of the polarized sea quark distributions from measurements 
of the Drell--Yan process. As discussed above, a measurement of the 
asymmetry as function of $x_F$ or $y$ can be directly inferred into 
the $x$-dependence of the polarized sea quark distributions.
The results obtained in this paper can be furthermore immediately
--~keeping in mind  the limitations discussed in the introduction~-- 
applied to the production of massive vector bosons in polarized 
proton--proton collisions 
at RHIC~\cite{rhic}. 

With the knowledge of the next-to-leading order corrections, it is 
furthermore possible 
to quantify the uncertainty of the theoretical prediction and hence the 
theoretical error on a measurement of the polarized sea quark distributions. 
We have demonstrated that the absolute value of the Drell--Yan asymmetry in 
the central region varies by less than 0.01 under variation of the mass 
factorization scale, demonstrating the perturbative stability of the 
prediction. This uncertainty is far smaller than the deviation between 
predictions obtained for different parameterizations of the polarized 
sea quark distributions.

\begin{appendix}
\setcounter{equation}{0}
\renewcommand{\theequation}{\mbox{\Alph{section}.\arabic{equation}}}
\section{Identities for ``$+$''--functions}
The $\e$--expansion of the singular terms in the parton level cross sections
of Section~\ref{sec:calc} yields
\begin{eqnarray}
(1-z)^{-1-\e} & = & -\frac{1}{\e}\delta (1-z) + \frac{1}{(1-z)_{+}} - 
\e\left(\frac{\ln (1-z)}{1-z}\right)_{+}, \nonumber \\
z_1^{-1-\e} & = & -\frac{1}{\e}\delta (z_1) + \frac{1}{(z_1)_{+}} - 
\e\left(\frac{\ln z_1}{z_1}\right)_{+}, \nonumber \\  
(1-z_1)^{-1-\e} & = & -\frac{1}{\e}\delta (1-z_1) + \frac{1}{(1-z_1)_{+}} - 
\e\left(\frac{\ln (1-z_1)}{1-z_1}\right)_{+}. \label{eq:plusid}
\end{eqnarray}
Integration over $z_1$
yields the following ``+''--functions, all defined on 
the interval $[0;1]$:
\begin{displaymath}
\frac{1}{(z_1^{*})_{+}},\qquad\frac{1}{(1-z^{*}_1)_{+}},
\end{displaymath}
and $\delta$-functions:
\begin{displaymath}
\delta(z_1^{*}),\qquad\delta(1-z_1^{*}).
\end{displaymath}
Furthermore, various products of the above terms appear.
By construction, we have $z=(\xaa\xbb)/(\xa\xb)$ while
$z_1^{*}$ is fixed by the definition of $x_F$ or $y$ respectively.
As the parton level cross section is always integrated over the parton 
momentum fractions $\xa$ and $\xb$, it is desirable to re-express 
the above functions in terms of
\begin{displaymath}
\delta(\xa-\xaa), \qquad \delta(\xb-\xbb)
\end{displaymath}
and
\begin{eqnarray}
\frac{1}{(x_i-x_i^0)_{+}} & \mbox{with} & 
\int_{x_i^0}^1 \d x_i \frac{f(x_i)}{(x_i-x_i^0)_{+}}\equiv
\int_{x_i^0}^1 \d x_i \frac{f(x_i)-f(x_i^0)}{x_i-x_i^0}, \nonumber \\
\left(\frac{\ln(1-x_i^0/x_i)}{x_i-x_i^0}\right)_{+} & \mbox{with} & 
\int_{x_i^0}^1 \d x_i f(x_i) 
\left(\frac{\ln(1-x_i^0/x_i)}{x_i-x_i^0}\right)_{+}
\nonumber \\ &&
\equiv
\int_{x_i^0}^1 \d x_i \left[f(x_i)-f(x_i^0)\right]  
\frac{\ln(1-x_i^0/x_i)}{x_i-x_i^0},  \\
\frac{1}{[(\xa-\xaa)(\xb-\xbb)]_{+}} & \mbox{with} & \int_{\xaa}^{1} \d \xa
\int_{\xbb}^{1} \d \xb \frac{f(\xa,\xb)}{[(\xa-\xaa)(\xb-\xbb)]_{+}}
\nonumber \\ &&
\equiv
\int_{\xaa}^{1} \d \xa\int_{\xbb}^{1} \d \xb
\frac{f(\xa,\xb)-f(\xaa,\xb)-f(\xa,\xbb)+f(\xaa,\xbb)}{(\xa-\xaa)(\xb-\xbb)} .
\nonumber
\end{eqnarray}

We first consider the $x_F$-distributions. 
All terms containing $\delta(1-z)$ can be simplified by
\begin{equation}
\delta(1-z) \, 
\delta\left(x_F-\left[z_1(\xa+\xb)(1-z)+z\xa-\xb \right]\right)
= \frac{\xaa \xbb}{\xaa + \xbb}\, \delta (\xa-\xaa)\delta(\xb-\xbb)
\label{eq:delx1}
\end{equation} 
and subsequent integration over $z_1$. For the remaining terms, we integrate 
over $z_1$ using the $x_F$-fixing $\delta$-function, which yields
\begin{displaymath}
z_1^{*}=\frac{\xaa+\xb}{(1-z)\xb(\xa+\xb)}\,(\xb-\xbb)\qquad\mbox{and}
\qquad
1-z_1^{*}=\frac{\xa+\xbb}{(1-z)\xa(\xa+\xb)}\,(\xa-\xaa).
\end{displaymath}

As the integrations over the parton momenta $\xa$ and $\xb$ are independent 
from each other, we can perform them in arbitrary order. For terms containing 
$\delta(z_1^{*})$ and $1/(z_1^{*})_{+}$ only, 
we choose to integrate $\xb$ first. For terms containing $\delta(1-z_1^*)$ and
 $1/(1-z_1^{*})_{+}$ only, $\xa$ is integrated first.
We can then make the following replacements:
\begin{eqnarray}
\delta(z_1^{*}) &=& \frac{(1-z)\xbb(\xa+\xbb)}{\xaa+\xbb}\,\delta(\xb-\xbb),
\nonumber \\
\frac{1}{(z_1^{*})_{+}} &=& \frac{(1-z)\xb(\xa+\xb)}{\xaa+\xb} \left[
\frac{1}{(\xb-\xbb)_{+}} + \delta (\xb-\xbb) \ln \frac{(\xaa+\xbb)(1-\xbb) 
\xa}{\xbb(\xa+\xbb)(\xa-\xaa)}\right] , \nonumber \\
\delta(1-z_1^{*}) &=& \frac{(1-z)\xaa(\xaa+\xb)}{\xaa+\xbb}\,\delta(\xa-\xaa),
\label{eq:delx2} \\
\frac{1}{(1-z_1^{*})_{+}} &=& \frac{(1-z)\xa(\xa+\xb)}{\xa+\xbb} \left[
\frac{1}{(\xa-\xaa)_{+}} + \delta (\xa-\xaa) \ln \frac{(\xaa+\xbb)(1-\xaa) 
\xb}{\xaa(\xaa+\xb)(\xb-\xbb)}\right]. \nonumber 
\end{eqnarray}
The above identities for $\delta(z_1^{*})$ and $\delta(1-z_1^{*})$ can be 
safely applied to terms containing ``$+$''--functions 
in $(1-z)$. These can then 
be rewritten as
\begin{eqnarray}
\delta(\xa-\xaa)\, \frac{1}{(1-z)_{+}} &=& 
\delta(\xa-\xaa) \left[ \frac{\xb}{(\xb-\xbb)_{+}} + \xbb\ln 
\frac{1-\xbb}{\xbb} \delta (\xb-\xbb)\right], \nonumber  \\
\delta(\xa-\xaa) \left(\frac{\ln(1-z)}{1-z}\right)_{+} &=&
 \delta(\xa-\xaa) \Bigg[ \xb \left(\frac{\ln (1-\xbb/\xb) }{\xb-\xbb}
\right)_{+} \label{eq:delx3}  \\
& & \hspace{1.5cm}
+ \xbb \left( \Li_2(\xbb) - \frac{\pi^2}{6} + \frac{1}{2} 
\ln^2 (1-\xbb) \right) \delta (\xb-\xbb) \Bigg]\nonumber
\end{eqnarray}
and similar for $(1\leftrightarrow 2)$.
These identities are however inapplicable to the only remaining term 
\begin{displaymath}
\frac{1}{(1-z)_{+}} \left[ \frac{1}{(z_1^{*})_{+}} + \frac{1}{(1-z_1^{*})_{+}}
\right] \frac{1}{(1-z)(\xa+\xb)},
\end{displaymath}
as  $\ln (\xa-\xaa)$ and $\ln (\xb-\xbb)$ appearing in (\ref{eq:delx2})
are not defined for $z=1$.
To overcome this problem, we rewrite
\begin{displaymath}
\ln (\xa-\xaa) = \ln \frac{\xa-\xaa\xbb}{\xbb} - \int_{\xbb}^{1} \d \alpha_2
\frac{1}{\alpha_2-\xaa\xbb/\xa},
\end{displaymath}
and similar for $(1\leftrightarrow 2)$. After subsequent algebraic 
manipulations and integration over $\alpha_1$ and $\alpha_2$, we obtain:
\begin{eqnarray}
\lefteqn{
\frac{1}{(1-z)_{+}} \left[ \frac{1}{(z_1^{*})_{+}} + \frac{1}{(1-z_1^{*})_{+}}
\right] \frac{1}{(1-z)(\xa+\xb)}=} \nonumber \\
& & \hspace{0.4cm}
\delta (\xa-\xaa) \delta (\xb-\xbb) \frac{\xaa \xbb}{\xaa +\xbb}
\Bigg[ \ln \frac{\xaa}{1-\xaa} \ln 
\frac{\xbb}{1-\xbb} - \Li_2(\xaa) -\frac{1}{2} \ln^2 (1-\xaa) \nonumber \\
& & \hspace{1.5cm}
- \Li_2(\xbb) -\frac{1}{2} \ln^2 (1-\xbb) + \frac{\pi^2}{2} \Bigg] \nonumber \\
& & +\; \delta (\xa-\xaa) \frac{\xaa\xb}{\xaa+\xbb} \Bigg[ \frac{1}
{(\xb-\xbb)_{+}} \ln \frac{(\xaa+\xbb)(1-\xaa)}{\xaa(\xaa+\xb)}-  
\left(\frac{\ln (1-\xbb/\xb)}{\xb-\xbb}\right)_{+} \Bigg] \nonumber \\
& & +\; \delta (\xb-\xbb) \frac{\xa\xbb}{\xaa+\xbb} \Bigg[ \frac{1}
{(\xa-\xaa)_{+}} \ln \frac{(\xaa+\xbb)(1-\xbb)}{\xbb(\xa+\xbb)} -
\left(\frac{\ln (1-\xaa/\xa)}{\xa-\xaa}\right)_{+} \Bigg] \nonumber \\
&& +\; \frac{(\xa + \xb) \xa\xb}{(\xaa+\xb) (\xa+\xbb) } \,
\frac{1}{[(\xa-\xaa)(\xb-\xbb)]_{+}} \label{eq:delx4}
\end{eqnarray}

Analogous relations can be obtained for the $y$-distributions, where we 
can first make the replacement
\begin{equation}
\delta (1-z) \, 
\delta \left(y - \frac{1}{2}\left[ \ln \frac{\xa}{\xb} + \ln 
\frac{z+z_1-zz_1}{1-z_1+zz_1} \right]\right)
= \xaa \xbb \delta(\xa-\xaa)\delta(\xb-\xbb).
\label{eq:dely1}
\end{equation}
Evaluation of the $z_1$--integral in the remaining terms yields
\begin{displaymath}
z_1^{*}=\frac{\xaa(\xb+\xbb)}{\xb(1-z)(\xa\xbb+\xaa\xb)}
\,(\xb-\xbb)\qquad\mbox{and}
\qquad
1-z_1^{*}=\frac{\xbb(\xa+\xaa)}{\xa(1-z)(\xa\xbb+\xaa\xb)}\,(\xa-\xaa).
\end{displaymath}
Only the identities (\ref{eq:delx3}) are unchanged for the 
$y$-distributions, the remaining identities read:
\begin{eqnarray}
\delta(z_1^{*}) &=& \frac{(1-z)(1+z)\xbb}{2z}\,\delta(\xb-\xbb),
\nonumber \\
\frac{1}{(z_1^{*})_{+}} &=& \frac{(1-z)\xb(\xa\xbb+\xaa\xb)}{\xaa(\xb+\xbb)} 
\left[
\frac{1}{(\xb-\xbb)_{+}} + \delta (\xb-\xbb) \ln \frac{2\xa\xaa(1-\xbb)}
{\xbb(\xa+\xaa)(\xa-\xaa)}\right] , \nonumber \\
\delta(1-z_1^{*}) &=& \frac{(1-z)(1+z)\xaa}{2z}\,\delta(\xa-\xaa),
\label{eq:dely2} \\
\frac{1}{(1-z_1^{*})_{+}} &=& 
\frac{(1-z)\xa(\xa\xbb+\xaa\xb)}{\xbb(\xa+\xaa)} 
\left[
\frac{1}{(\xa-\xaa)_{+}} + \delta (\xa-\xaa) \ln \frac{2\xb\xbb(1-\xaa)}
{\xaa(\xb+\xbb)(\xb-\xbb)}\right],  \nonumber
\end{eqnarray}
and
\begin{eqnarray}
\lefteqn{
\frac{1}{(1-z)_{+}} \left[ \frac{1}{(z_1^{*})_{+}} + \frac{1}{(1-z_1^{*})_{+}}
\right] \frac{1}{1-z}=} \nonumber \\
& & \hspace{0.4cm}
\delta (\xa-\xaa) \delta (\xb-\xbb) \xaa \xbb
\Bigg[ \ln \frac{\xaa}{1-\xaa} \ln 
\frac{\xbb}{1-\xbb} - \Li_2(\xaa) -\frac{1}{2} \ln^2 (1-\xaa) \nonumber \\
& & \hspace{1.5cm}
- \Li_2(\xbb) -\frac{1}{2} \ln^2 (1-\xbb) + \frac{\pi^2}{2} \Bigg] \nonumber \\
& & +\; \delta (\xa-\xaa) \frac{\xaa\xb(\xb+\xbb)}{2\xbb} \Bigg[ \frac{1}
{(\xb-\xbb)_{+}} \ln \frac{2\xbb(1-\xaa)}{\xaa(\xb+\xbb)}-  
\left(\frac{\ln (1-\xbb/\xb)}{\xb-\xbb}\right)_{+} \Bigg] \nonumber \\
& & +\; \delta (\xb-\xbb) \frac{\xa\xbb(\xa+\xaa)}{2\xaa} \Bigg[ \frac{1}
{(\xa-\xaa)_{+}} \ln \frac{2\xaa(1-\xbb)}{\xbb(\xa+\xaa)} -
\left(\frac{\ln (1-\xaa/\xa)}{\xa-\xaa}\right)_{+} \Bigg] \nonumber \\
&& +\; \frac{(\xa\xbb+\xaa\xb)^2\xa\xb}{\xaa\xbb(\xa+\xaa)(\xb+\xbb)} \,
\frac{1}{[(\xa-\xaa)(\xb-\xbb)]_{+}} \label{eq:dely3}
\end{eqnarray}

\end{appendix}

\goodbreak

\end{document}